\documentclass{article}

\usepackage{arxiv}

\usepackage[utf8]{inputenc} 
\usepackage[T1]{fontenc}    
\usepackage{hyperref}       
\usepackage{url}            
\usepackage{booktabs}       
\usepackage{amsfonts}       
\usepackage{nicefrac}       
\usepackage{microtype}      
\usepackage{lipsum}		
\usepackage{graphicx}
\usepackage[minnames=2,maxnames=3,backend=bibtex, sorting=none]{biblatex}
\usepackage{doi}
\usepackage{amsmath}  
\usepackage{amsfonts} 
\usepackage{graphicx} 
\usepackage{comment}
\usepackage{scrextend}
\usepackage[dvipsnames]{xcolor}
\usepackage{footnote}
\usepackage{caption}
\usepackage{subcaption}
\usepackage{float}
\usepackage{multicol}
\newenvironment{Figure}
  {\par\medskip\noindent\minipage{\linewidth}}
  {\endminipage\par\medskip}

\begin{document}

\title{The Completely Hackable Amateur Radio Telescope (CHART) Project}

\author{Lindsay~M.~Berkhout$^{1}$\thanks{E-mail: lmberkhout@asu.edu},
Adam~P.~Beardsley$^{2}$,
Daniel C. Jacobs$^{1}$, Raven Braithwaite$^{1}$,\\ 
\textbf{Bryanna Gutierrez-Coatney$^{1}$, Arib Islam$^{1}$, Ahlea Wright$^{2}$}\\
$^{1}$School of Earth and Space Exploration, Arizona State University, Tempe, AZ, USA\\
$^{2}$Department of Physics, Winona State University, Winona, MN, USA}

\date{}


\title{The Completely Hackable Amateur Radio Telescope (CHART) Project}
\maketitle 

\begin{abstract}
We present the Completely Hackable Amateur Radio Telescope (CHART), a project that 
provides hands-on radio instrumentation and design experience to undergraduates
while bringing accessible radio astronomy experiments to high school students and teachers.
Here we describe a system which can detect 21-cm emission from the Milky Way which is optimized for cost and simplicity of construction. Software, documentation, and tutorials are all completely open source to improve the user experience and facilitate community involvement.
We demonstrate the design with several observations which we compare with state-of-the-art surveys.
The system is shown to detect galactic 21-cm emission in both rural and urban settings.
\end{abstract}
\begin{multicols}{2}

\section{Introduction} 

The Completely Hackable Amateur Radio Telescope (CHART) Project 
provides a platform and tutorials for amateur radio astronomy, with the intent of broadening access to radio science at the secondary school and early undergraduate level. 

A radio telescope is an excellent educational or amateur astronomy project for several reasons. Optical astronomy is popularized by high profile telescopes such as the Hubble Space Telescope or the James Webb Space Telescope \cite{jwst} and there are multiple paths for an interested amateur to obtain their own backyard optical telescope. However, optical observations are best done under clear skies from dark rural locations, and most people live in cities where air quality can be low or clouds are common. About 80\% of North Americans cannot see the Milky Way from their homes \cite{pollution}. On the other hand, while cities also pose noise challenges, with careful design it is possible to see the Milky Way at radio frequencies from even a large city.  Additionally, radio frequency observations can illuminate different properties of astronomical objects that cannot be observed with low-cost optical instruments, such as Doppler shift from motion and spectral properties of galaxies.  Lastly, in a benefit particularly useful to k-12 schools, radio observations can be performed in the day time when class is in session.

Though several previous projects have described amateur-grade systems using consumer grade electronics, the  part selection, signal processing, and software analysis details are usually left to the user. This increases barriers to participation for those with less technical experience. 
Here we describe a radio telescope kit which, through a combination of documentation and testing, can be built by a typical high school science teacher or someone with similar experience.

The base platform targets measurements of the 21-cm line of neutral hydrogen. Low cost, easy to obtain materials are used for the design, and the open source system design and software are available online\footnote{\url{https://https://astrochart.github.io/}}. The total cost currently runs at about \$300 of materials, with efforts being made to reduce this further. The design is ``hackable" in the sense that enough documentation is provided for users to make improvements or branch out in new directions. The project is also educational for undergraduate students at participating universities who do most of the development work. This contributes further to CHART's educational goals helping the next generation of astronomers get hands-on experience with instruments as well as improves representation at the interface for kit users.

Here we demonstrate CHART with a short primer on our object of choice (Sec.~\ref{sec:science}), a description of the hardware and software design (Sec.~\ref{sysdesign}), and example observing sessions in multiple settings (Sec.~\ref{results}). We end with a short discussion on the classroom use of CHART (Sec.~\ref{class}).

\section{Observing the 21\,cm Line}\label{sec:science}
Atomic neutral hydrogen emits a 
spectral line when the electron transitions between two hyper-fine levels of the ground state. 
This ``spin-flip" transition occurs when the proton and electron spins go from being aligned to anti-aligned, emitting a photon with rest wavelength 21\,cm, or frequency 1420.4\,MHz.
For neutral hydrogen clouds in the Milky Way and nearby galaxies, an observed deviation from this frequency is a Doppler shift caused by motion, and can be mapped directly to a velocity for the observed gas. 

The electromagnetic energy at this particular part of the spectrum can easily penetrate through cosmic dust and Earth's atmosphere, making it an easy target for ground observations. 
1420\,MHz is in a protected frequency band for radio astronomy, encompassing 1400-1427 MHz in the United States\footnote{\url{https://www.ntia.doc.gov/files/ntia/publications/2003-allochrt.pdf}}, and therefore should not be subject to radio frequency interference (RFI) from other sources.  
For these reasons, radio astronomy with the 21\,cm line is an ideal tool for students to learn about emission spectra, galactic motion, or even evidence for dark matter with rotation curves \cite{Levine_2008}.

Galactic hydrogen has been observed many times with many instruments, beginning with Ewen \& Purcell in 1951\cite{1951Natur.168..356E}, and followed by a number of more recent surveys \cite[e.g.][]{smoot, salsamap,surveycomp}. Recently hobbyist interest in 21cm radio science has picked up, and there are many avenues to participate at the amateur level, such as the Society for Amateur Radio Astromomy (SARA) grants \footnote{\url{https://www.radio-astronomy.org/grants}}, the Goldstone Apple Valley Radio Telescope \cite{gavrt}, and the SALSA project \cite{salsa}, but these opportunities often either assume pre-existing radio science literacy or provide access to a telescope for use, where one does not design or build their own instrument. There is still a significant gap to bridge between the amateur radio astronomy community and high school level physics and astronomy.

Regarding 21-cm amateur measurements specifically, there are a number of projects with similar goals. Each project takes a unique approach with differing goals and learning outcomes, and we include a few of the most similar works here. The Digital Signal Processing in Radio Astromomy (DSPIRA) \footnote{\url{https://wvurail.org/lightwork/}} project focuses on teaching signal processing and fourier analysis in this regime. The Physics Open Lab website reports measurements of the Milky way made with an amateur telescope \footnote{\url{https://physicsopenlab.org/2020/09/08/milky-way-structure-detected-with-the-21-cm-neutral-hydrogen-emission/}}. The PICTOR telescope \footnote{\url{https://pictortelescope.com/}} offers measurements of the Milky Way from a remote setup.  The BHARAT \cite{bharat} telescope uses a similar off-the-shelf principle to construct an amateur telescope for use in undergraduate labs.

The CHART project aims to build on these initiatives by targeting its content towards secondary school teachers as an audience, and providing detailed, open source materials and code. The project is organized as a "follow along" set of tutorials taking a user from building their own horn to a rotation curve constructed with their data.
We use accessible materials for building a low-cost instrument and expect no pre-existing radio science expertise.

\begin{Figure}
    \centering
    \includegraphics[height=\columnwidth, angle=270]{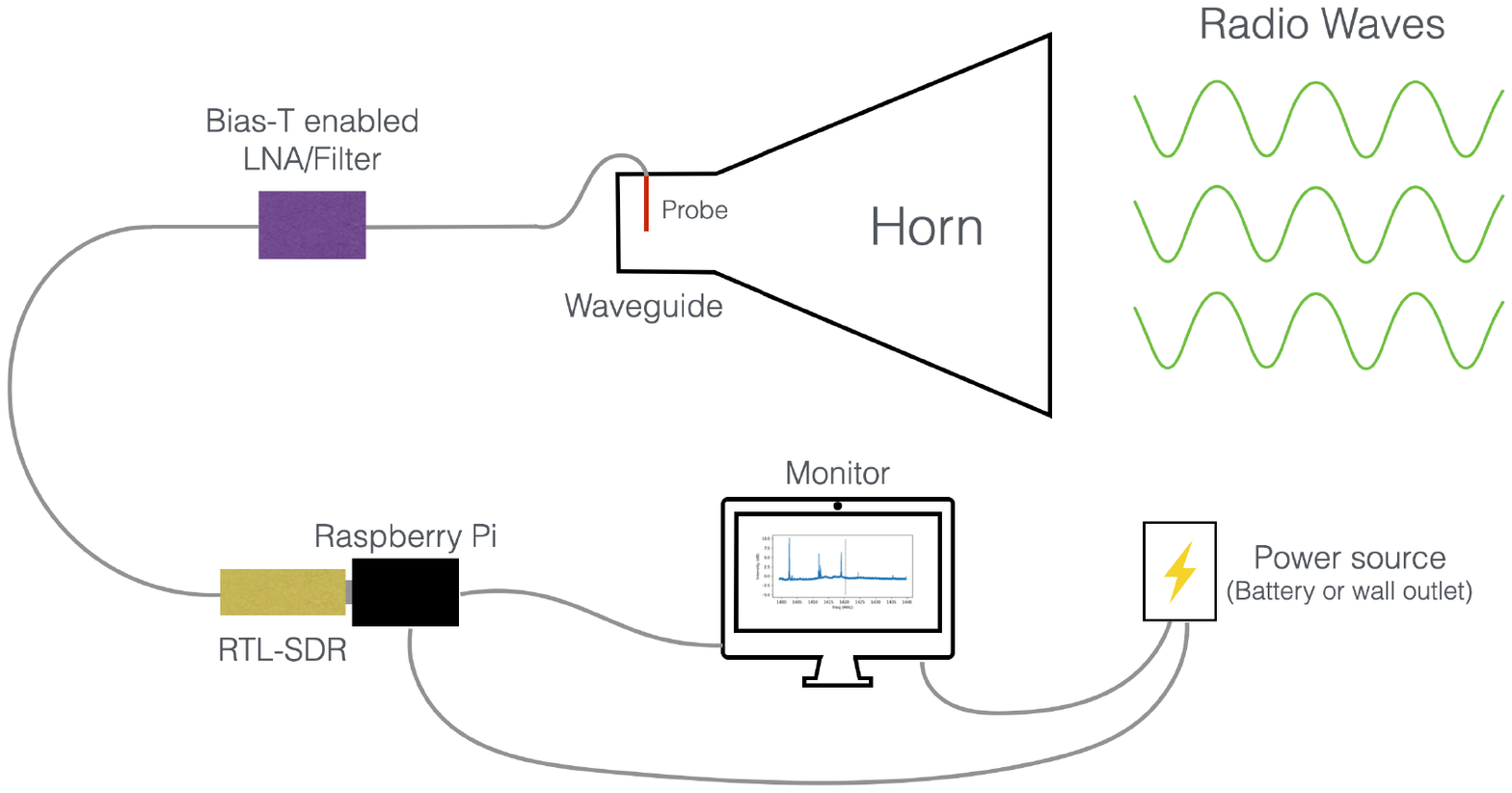}
    \captionof{figure}{The full system diagram for a CHART setup. Electromagnetic radiation enters the horn, and is picked up by the probe inside the waveguide. The probe connects to a combination noise amplifier (LNA) and filter via coaxial cable. The LNA/filter connects to an RTL-SDR software defined radio which is read out by a Raspberry-Pi processor. The RTL provides power to the LNA as a DC bias on the RF cable. A monitor is needed to view the Pi interface, and battery power is needed for both the Pi and monitor.}
    \label{fig:system}
\end{Figure}

\section{System Design} \label{sysdesign}
A summary of a fully tested example system is included in figure \ref{fig:system}. Due to the intended reconfigurability of CHART, many substitute components and designs could be used. This iteration is suggested as an easy, low-cost starting point, and has been well tested by the project participants. The design has been optimized to minimize part count which improves portability and ease of setup.

\subsection{Feed-Horn and Antenna}
The design of the CHART front-end was optimized for observations at the rest frequency of neutral hydrogen, as well as cost-effectiveness and ease of construction. A feed-horn and antenna configuration was chosen as it can be easily constructed from aluminium wrapped cardboard and a length of wire.

The purpose of the feed horn is to provide directional gain, acting as a "funnel" for the desired radiation. The dimensions for the horn were chosen using electromagnetic simulations.
The optimized parameter was the beam directivity, or the concentration of an antenna's radiation pattern in a particular direction, at 1420 MHz. 

Figure \ref{fig:dimensions} shows the dimensions of the horn used for the measurements in this paper.

A wire antenna of length 6.3\,cm is soldered into a coaxial connector and installed in the side of the waveguide (the rectangular portion at the bottom of the horn), using a soup can lid as a structural support and grounding point.

A photo of the fully constructed front end is included in figure \ref{fig:CHART}. 

\begin{Figure}
    \centering
    \includegraphics[height=\columnwidth, angle=270]{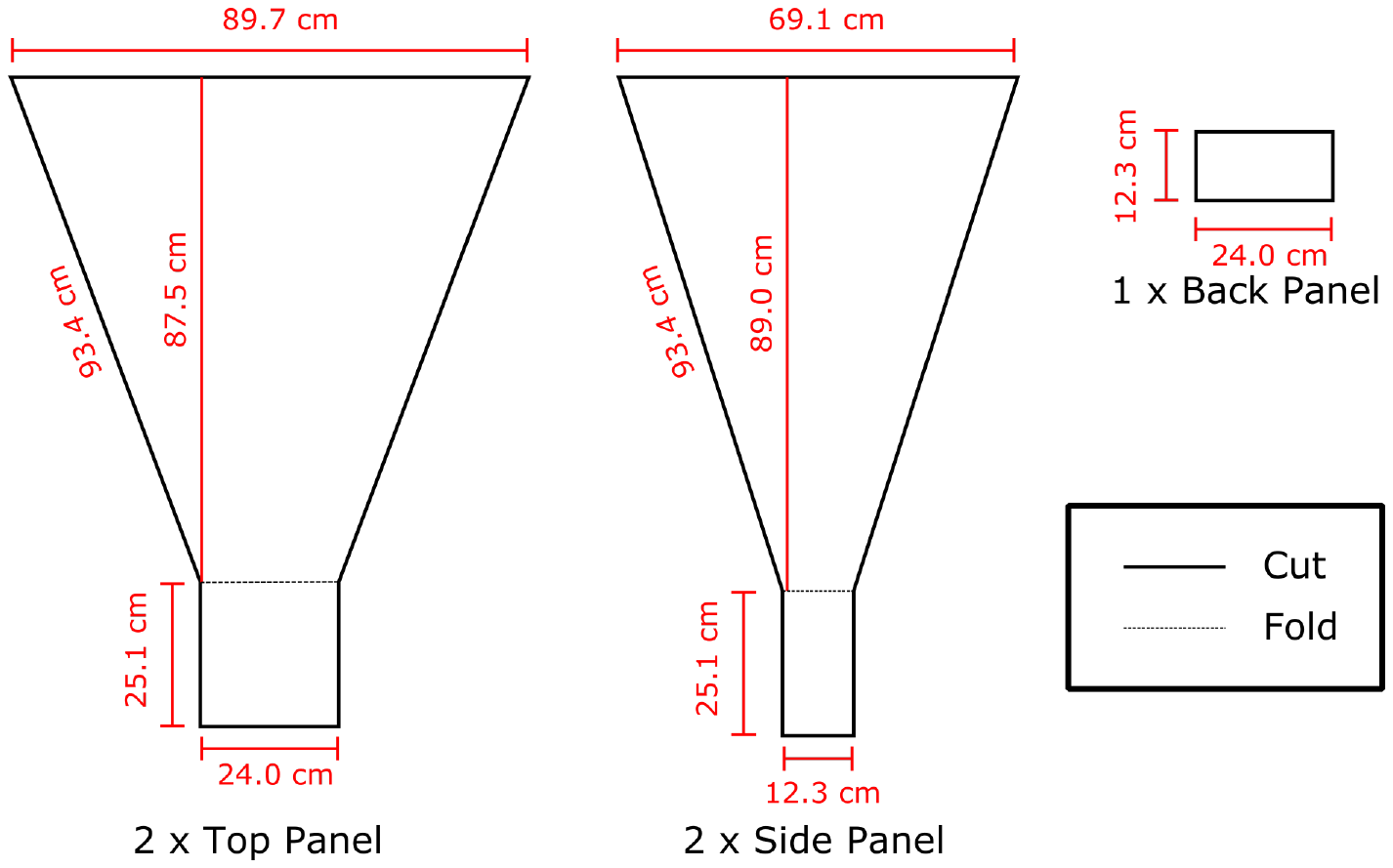}
    \captionof{figure}[width=\columnwidth]{The dimensions and template for the CHART horn construction.}
    \label{fig:dimensions}
\end{Figure}

\subsection{Electronics}
The analog signal chain filters out unwanted signals and amplifies the radiation picked up at the antenna. 

A few different amplification and filtering schemes have been tested. The current suggested hardware based on performance, cost, and ease of use is a Nooelec brand combination bias-tee enabled low noise amplifier (LNA) and filter. The Nooelec module combines both the filtering and amplification into one component and is easily obtained for minimal cost from consumer outlet targeted sellers, and combines both the filtering and amplification into one component. 

Additionally, the module can be powered using a bias-tee, meaning it can be powered by a USB radio over the RF coaxial cable instead of needing a separate power source. 

The module has a good performance with about 40 dB of gain over 1375-1450 MHz and a noise figure of 1.05dB or 79K. The attenuation outside of the 65MHz pass-band is between 40 and 60dB.

\begin{Figure}
    \centering
    \includegraphics[width=\columnwidth]{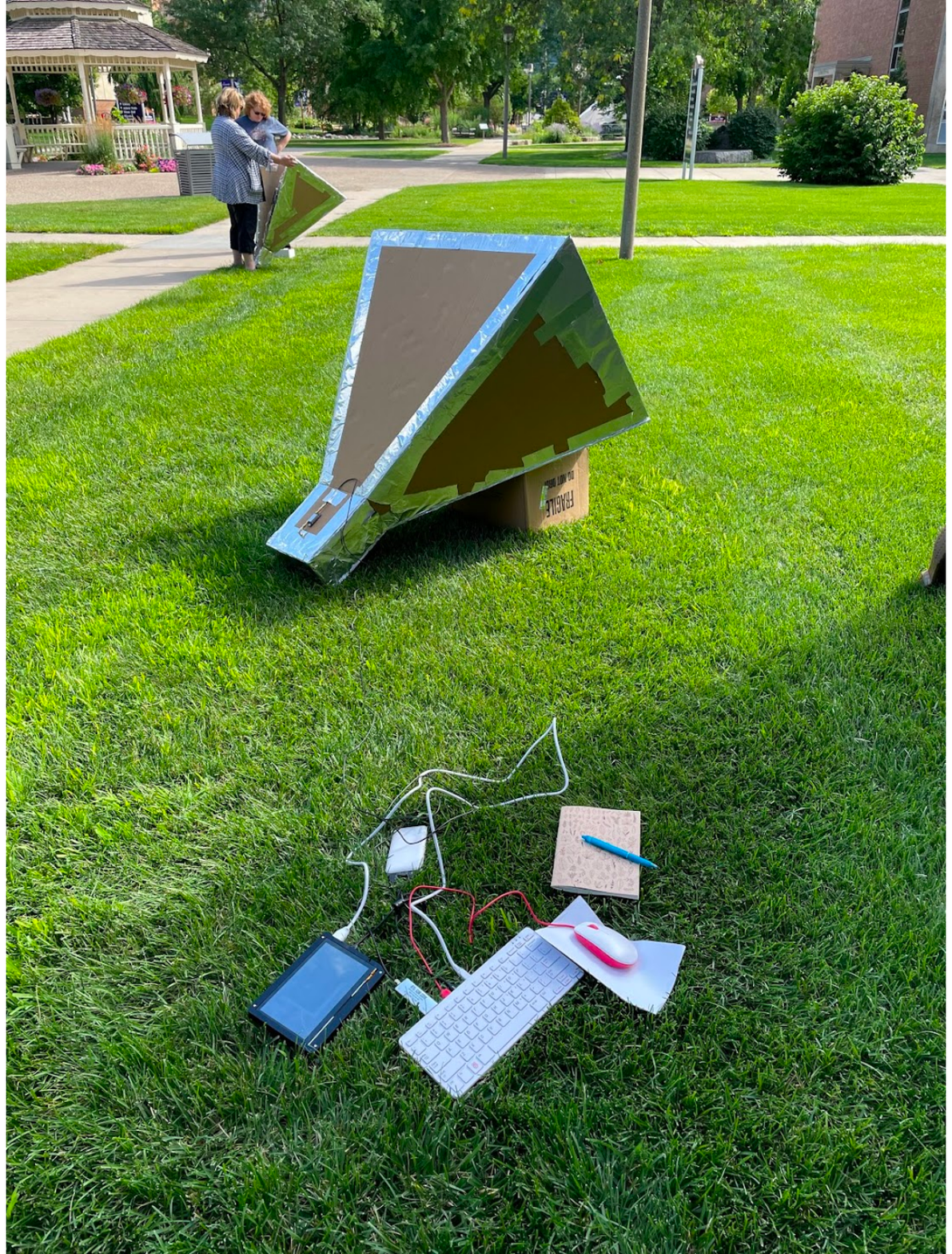}
    \captionof{figure}{A fully constructed CHART horn on an observing trip at a weeklong workshop for High School teachers at Winona State University. The CHART documentation kit includes instructions for making a pyramidal horn made from cardboard and tin foil, as well as for setting up the electronics and software.}
    \label{fig:CHART}
\end{Figure}

\subsection{Mixing and Digitization}
The output of the amplified and filtered analog signal is mixed down to a lower frequency and digitized. The digital voltage samples are read out by a computer, transformed into a spectrum, and averaged. 

Here we have used the "RTL-SDR Blog V3," a popular hobbyist software defined radio which is based on the RTL2832U and Rafael Micro R820T2 chipsets. The RTL-SDR model is a USB dongle that can be obtained for approximately \$30. It includes a tunable mixer, adjustable gain blocks, digitizer, and bias-tee for powering an external amplifier. The frequency range of operation is 24 – 1766 MHz, although with direct sampling can go as low as 500 kHz, and the maximum sample rate (without dropped samples) is  2.56 MS/s. 

We use a Raspberry Pi to process the samples from the SDR, and our tutorials include simple instructions to install and configure the necessary software.
This setup provides easy to setup and low cost computing and file storage until the data can be transferred onto a personal computer or server, although any machine that can support a USB SDR and the GNURadio software described section \ref{sec:software} could be substituted for the Pi.

\subsection{Software} \label{sec:software}

The open source and free GNURadio\footnote{gnuradio.org} provides the basis of our data collection.
One can use GNURadio's graphical user interface (GUI) directly, building a signal processing flowchart to acquire a spectrum and write it to a file.
We have also written wrappers to streamline the data collection in our open source software package\footnote{https://github.com/astrochart} -- either in a Linux command line mode, or our dedicated GUI. 
This is simpler to run, but it does obscure the data taking flow from the user.

The data taking options allow for flexibility depending on the need and expertise of the user.
Those who work directly with GNURadio will interact with the engineering aspects of radio instrumentation and learn more signal processing.
For those who are more interested in the astronomical analysis of the data, the  CHART software provides a quicker method of obtaining data. 
A custom python package provides analysis functions. Functions include coordinate conversions, bandpass calibration, and plotting in a Doppler velocity frame.

\subsection{Tutorials}

To make the platform as user friendly as possible, and to provide usability for the widest range of experience levels, detailed tutorials are available on the CHART website. There are videos and text walkthroughs of the full system setup.

As this setup is reasonably complex and users are not expected to be familiar with advanced computing and coding, the tutorials also cover setting up the Raspberry Pi computer, basics of the Linux operating system, and use of the software. Analysis of the collected data is demonstrated in a Jupyter notebook. This notebook can be run directly on the Raspberry Pi, on any personal computer, or on a community server described below.

\subsection{Data Storage and Analysis Server}
We created a Microsoft Azure server to store data and perform analysis to facilitate community engagement, data sharing, and ease of use. For the pilot implementation, we allocated a virtual machine with 8 CPUs (3rd generation AMD Milan), 32 GB RAM, and a 128 GB hard drive with the option to expand as needed. This should comfortably meet the needs of about ten users.

Participants can contact us to acquire an account on the server. With an account, users can  upload their data directly from the CHART observer GUI. The server accepts the data upload and adds it to a database which includes metadata about each observation (e.g., location, date, observed frequencies). All uploaded data is visible to all users so they can easily exchange observations and compare results.

Once uploaded the users can look at their data in a JupyterHub\footnote{https://jupyterhub.readthedocs.io/} instance running on the server. 
User accounts are created with all necessary software for data analysis, including the CHART python package, and an example analysis Jupyter Notebook which users can use as a starting point and serves as a self-documented tutorial. This setup avoids initial setup difficulties which often pose a source of friction for new users.

\section{Example Observations} \label{results}
 While unwanted radio frequency interference (RFI) can be filtered out with a number of analog and digital techniques, interfering signals common in urban environments can be loud enough to cause distortion in the signal chain which can alter the astronomical signal irrecoverably. This makes doing radio astronomy in populated areas more difficult, but not impossible. 
 Results are presented here for two places with differing population densities. 
 This setup has been tested by students in near Winona State University in Winona, Minnesota and in downtown Phoenix near Arizona State University.

\begin{Figure}
    \centering
    \includegraphics[height = \columnwidth, angle=270]{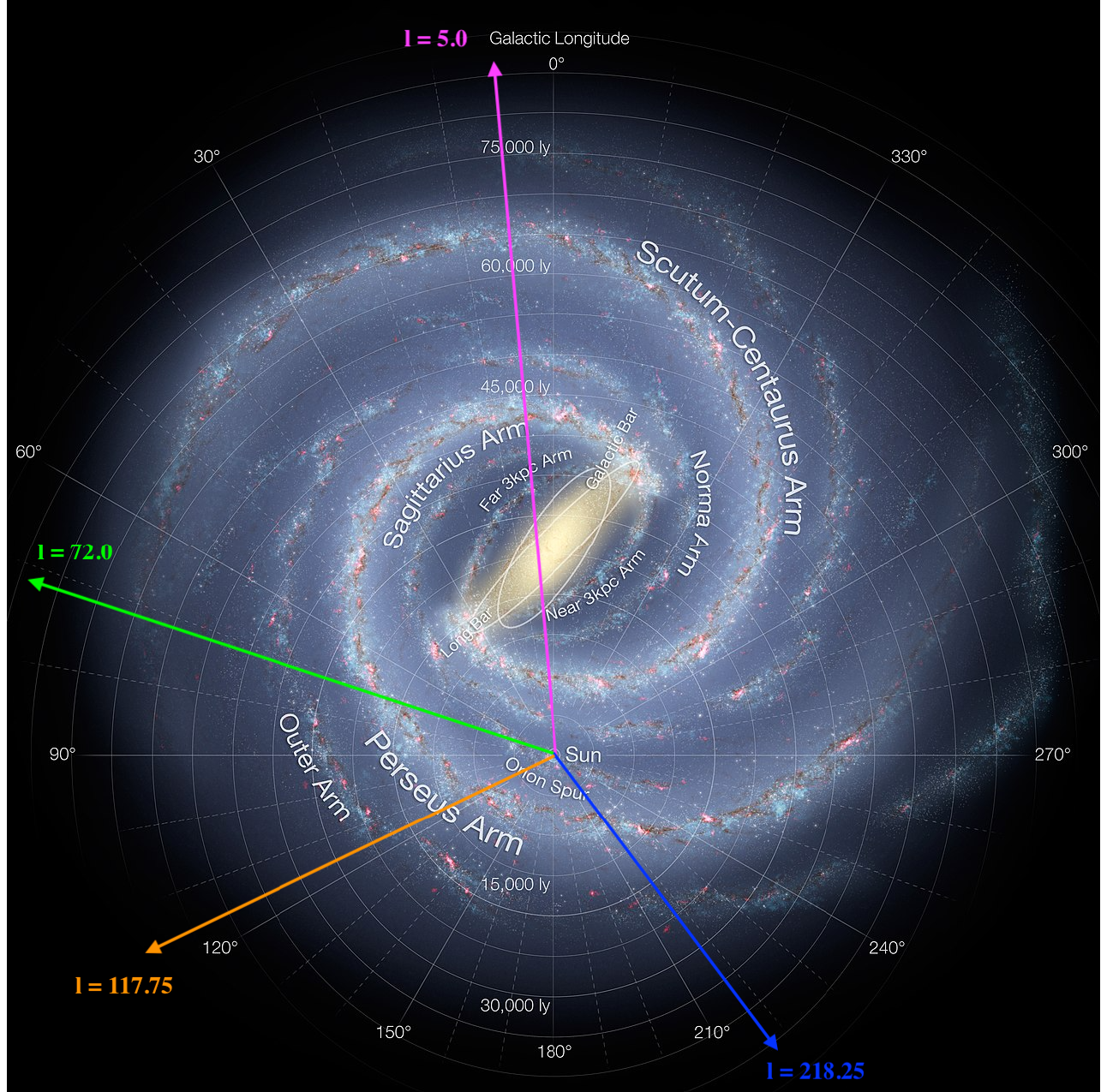}
    \captionof{figure}{An artist’s concept showing principle structures of the Milky Way and illustrating observed sight lines. The labels of the sight lines correspond to the locations of our observations in sections \ref{results1} and \ref{results2}. Image credit: NASA/JPL-Caltech/R. Hurt (SSC-Caltech) with this link: https://www.nasa.gov/jpl/charting-the-milky-way-from-the-inside-out.}
    \label{fig:milkyway}
\end{Figure}

\begin{Figure}
    \centering
    \includegraphics[height=\columnwidth]{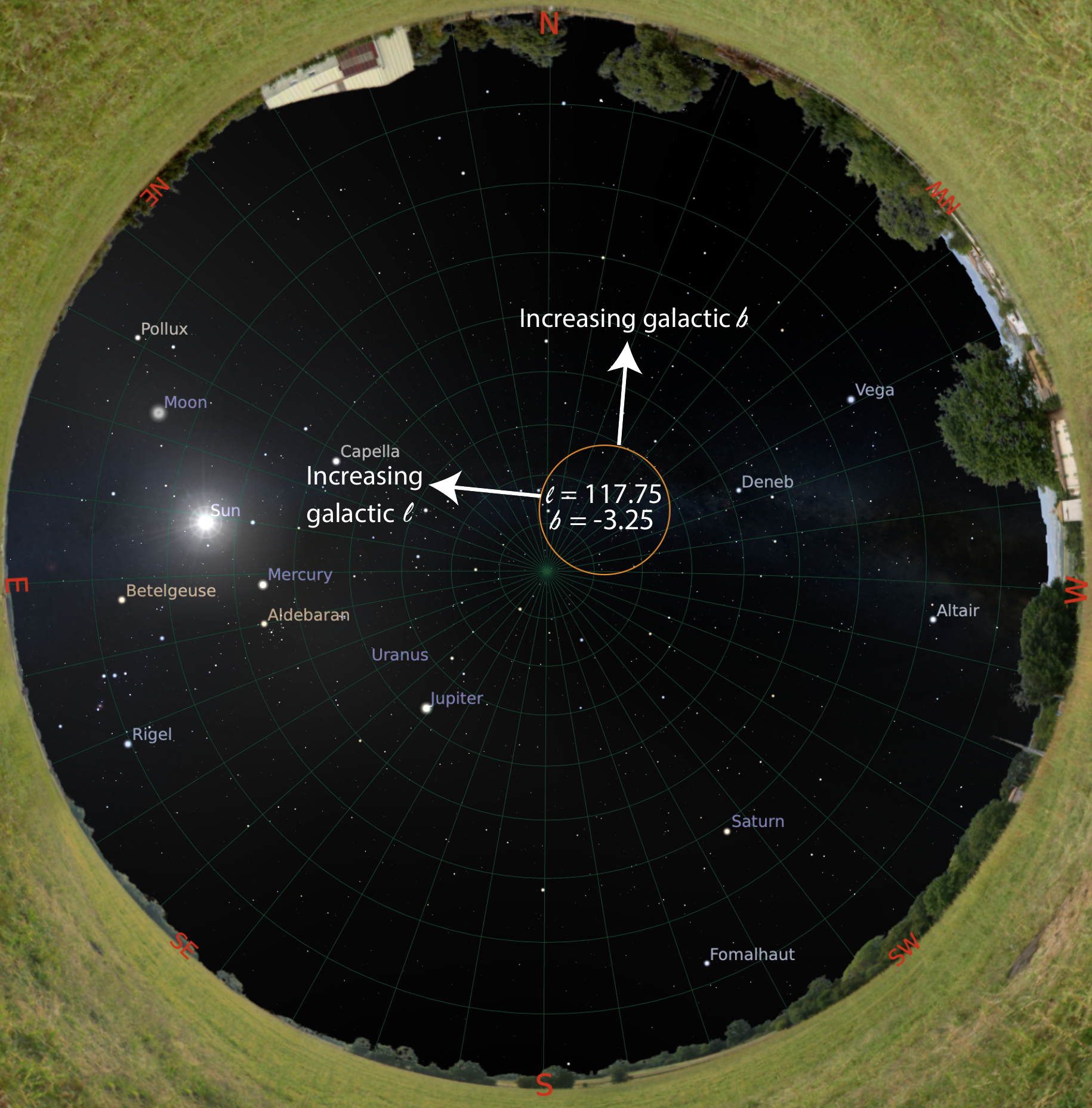}
    \captionof{figure}{A user's view from the Stellarium web simulator of the sky above the Winona, Minnesota observing location at the time of data collection. 
    The circle corresponds to the center of the pointing labelled as $l=117.75$ in figure \ref{fig:milkyway}. The size of the circled area approximates the CHART beam size.}
    \label{fig:stell}
\end{Figure}

\subsection{Methods} \label{methods}
Observations were conducted by several groups of undergraduate students. 

In each observation the time and orientation were noted. Neutral gas in the Milky Way is distributed mostly within a thin disk that is co-located with the optically visible galactic plane. 
Because the solar system is located within this disk,
any line of sight across the sky will contain galactic HI \cite{HIreview}. In most cases orientations were chosen towards the galactic plane to maximize detection probability, as the signal is strongest there. Observations included a wide scan across 20 MHz (2\,MHz at a time due to the limits of the SDR) surrounding the 21cm line to assess the interference environment. 

The center of the telescope pointing was determined by estimating the right ascension and declination of the horn pointing from measured azimuth and elevation. This was then converted to galactic coordinates, and the pointings will be labelled as such. In galactic coordinates, $l$ is the galactic longitude and $b$ is galactic latitude; degrees away from the galactic plane. Figure \ref{fig:milkyway} shows the line of sight for the galactic longitudes ($l$) used in our analysis. An example view on the sky of one of the pointings shown in figure \ref{fig:milkyway} is included in figure \ref{fig:stell}. This snapshot is from the Stellarium software\footnote{\url{https://www.stellarium.org}}, and approximates the area on the sky and location encompassed by one of our pointings. All measurements were taken with the horn described in section \ref{sysdesign}, which has a beam width of approximately 25 degrees.

Post observation, the data must be preprocessed for analysis. Many of our analysis choices were motivated by desire to compare our results to known neutral hydrogen profiles from the Milky Way. We chose to compare to the Leiden/Argentine/Bonn (LAB) survey \cite{surveycomp}. The LAB survey used the Dutch 25m telescope in Dwingeloo and the Argentinean 30m telescope in Villa Elisa to map the neutral hydrogen in the Milky Way. 

In order to compare to the LAB survey directly, we take a few processing steps. We must remove effects imposed on the astronomical data by our CHART system to calibrate the data to brightness temperature units, as well as transform to a common reference frame of motion to match the survey reference frame choice. 

First, we remove the instrumental effects. The CHART system itself imposes some characteristics onto our true sky signal, namely a multiplicative gain (e.g. from the antenna and amplifiers) and an additive noise (predominantly thermal noise from electronics). Most components in our system have a gain and noise profile that is relatively constant over the frequencies of interest. The one notable exception is the bandpass of the anti-aliasing filter in the Software Defined Radio, which prevents unwanted frequencies from aliasing into the data at the digitizing state. However, the filter varies strongly as a function of frequency and imposes an undesirable shape onto the spectrum. We measure this bandpass shape using an "off tuning" away from the 21-cm line where we expect no astronomical signal, and divide it out of the data.

After the bandpass shape is removed from the data, we  assume that the other contributions to gain and noise are linear in frequency in a small range around the 21-cm line. Then, mathematically, we know how these terms relate to the true sky. A two component model for the analog system is described by Eq. \ref{modeleq} which relates the measured spectrum ($d(\nu)$) to the true sky ($m(\nu)$) via an unknown multiplicative gain ($g$) and an unknown additive noise level ($n$) that are constant across the spectrum.
\begin{equation} \label{modeleq}
    d(\nu) = g \cdot m(\nu) + n
\end{equation}

We calculate $g$ and $n$ for the CHART data by using a model for the true brightness temperature ($T_B$) of the 21cm sky ($m$) in units of Kelvin (K). For this true sky model, we use a data-based simulation of the expected 21\,cm spectrum from the EU-HOU project, using their web simulator \footnote{\url{https://www.astro.uni-bonn.de/hisurvey/euhou/LABprofile/index.php}}. The simulator uses data from the LAB survey and allows the user to input their telescope field of view and observing coordinates to estimate what their expected spectrum should look like. The simulator has a maximum beam width of 20 degrees, which is
smaller than the CHART horn beam by about 5 degrees,
but provides a reasonable approximation of the field of
view. 

First we estimate the noise level by choosing a frequency where the model predicts no 21cm emission and is free of interference, allowing us to approximate ($d(\nu)) \approx n$. Then we assume this noise is the same across all frequencies, so we can then calculate the gain term ($g$) where the model is largest ($g \approx (d(\nu) - n)/m(\nu)$). 

We then apply these parameters across the spectrum. 
This method turns the uncalibrated CHART data into brightness temperature units. Now the data are on the same scale as the model, and the profile shapes can be compared directly.

After we remove the effects imposed by our instrument, we must convert our data to the kinematic reference frame used by the LAB survey. The velocity at local standard of rest (VLSR) observing frame is used. The LSR follows the mean motion of material in the local Milky Way, defined as stars in radius 100 pc from the Sun \cite{physuniv}. This frame has two parts: the motion of the Sun relative to the LSR, and the orbital motion of the Earth. We use the `astropy` software package for this conversion \cite{astropy:2022}, and apply the computed doppler shift for our location, observation time, and line of sight to the frequency axis.

\begin{Figure}
    \centering
    \includegraphics[width=\columnwidth]{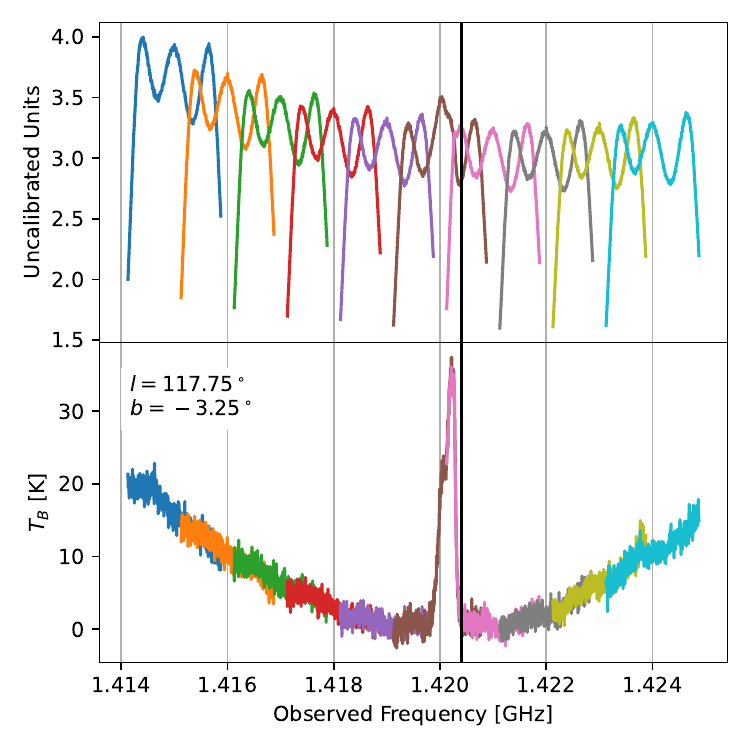}
    \captionof{figure}{\emph{Top:} The uncalibrated CHART data for galactic coordinates $l=117.75^\circ$ and $b=-3.25^\circ$. The individual spectra represent separate 2 MHz tunings of the SDR. The spectrometer scans across the frequency range in 2 MHz tunings, with a step of 1 MHz so that there is some overlap between tunings. The repeating structure in every tuning is due to the anti-aliasing filter of the SDR. 
    \emph{Bottom:} Calibrated data after dividing out the bandpass, subtracting noise, and correcting for the overall gain.}
    \label{fig:uncalibrated}
\end{Figure}

The raw, uncalibrated and uncorrected data for the sight line of $l=117.75^\circ$ taken in Winona, Minnesota is included in figure \ref{fig:uncalibrated} (top).  The shape of the spectrum is dominated by the anti-aliasing filter which is applied to each 2MHz tuning. This shape is divided out in the calibration post-processing step.
Sporadic RFI can be seen in a few tunings, and the 21cm line is visible slightly to the left of the rest frequency indicated as a bold vertical line. 
The whole frequency range has a slight slope caused by the uneven spectral response of the electronics. The vertical axis units are arbitrary and have no correspondence to a physical unit before calibration. Fig.~\ref{fig:uncalibrated} (bottom) shows the result of the calibration process. There is still an overall 'u' shape to the bandpass as we do not correct for second order effects when removing noise, but as the 21-cm emission occupies a small fraction of the bandwidth in the observation, we do not worry about removing effects outside of this range. 

\subsection{Results from Winona, Minnesota} \label{results1}

Figure \ref{fig:winona} shows the velocity profile for three different sets of galactic coordinates, following the corresponding lines of sight in figure \ref{fig:milkyway}. The data was taken near Winona State University in Winona, Minnesota. The CHART data is plotted against the LAB survey simulation for the same set of galactic coordinates. 

\begin{Figure}
    \centering
    \includegraphics[height=\columnwidth]{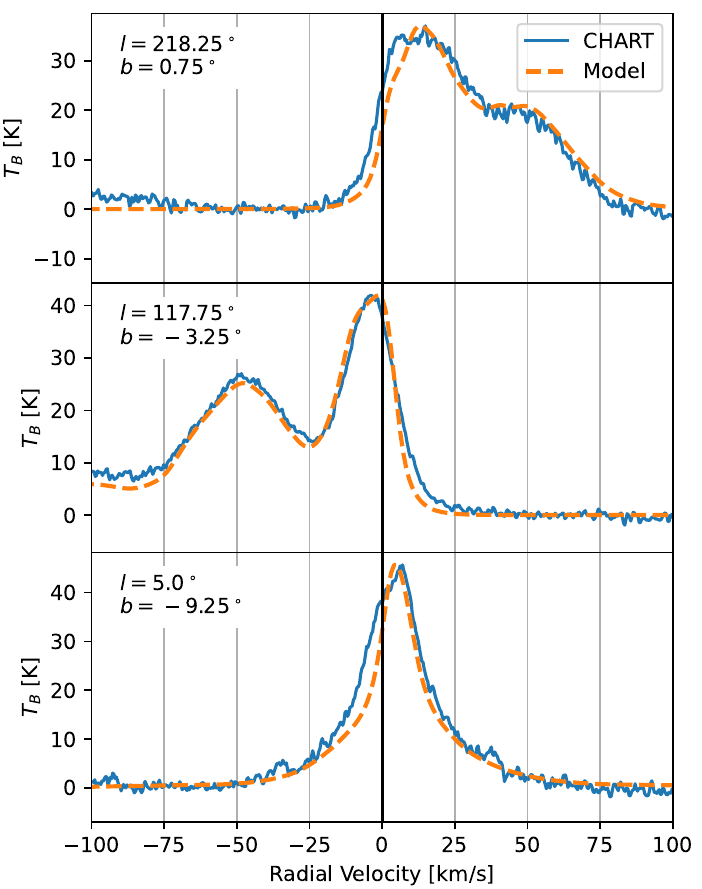}
    \captionof{figure}{Subfigures show a radial velocity vs brightness temperature measurement taken for 3 different sets of galactic coordinates, rounded to the nearest quarter degree. The coordinates follow the line of sight in figure \ref{fig:milkyway}. The data labelled "CHART"  correspond to the profile from our CHART telescope, and the "Model" is calculated with the EU-HOU web simulator.}
    \label{fig:winona}
\end{Figure}

Our data largely agree with the simulation results for the selected galactic coordinates. The locations of peaks in the spectra are well matched  between our data and the survey simulation data. There is minor variation in the shapes of the profiles, which could result from a variety of factors. Likely culprits are pointing accuracy errors or variation in the field of view of the data and models. Pointing accuracy is limited by the observer's ability to estimate the coordinates of the center of their pointing with a protractor and phone compass, and the beam size and shape is not exact between the EU-HOU simulation and the real horn. 

\subsection{Results from Phoenix, Arizona} \label{results2}
This section presents a more challenging RFI test environment than the previous test location. Measurements were made in near the city center in Phoenix, Arizona, and were taken in the day time.  In this location we expect interference at frequencies close to the HI band (including cell phones, FM radio, and other transmitters) at levels strong enough to overcome the bandpass filter and possibly to saturate the amplifier.

\begin{Figure}
    \centering
    \includegraphics[width=\columnwidth]{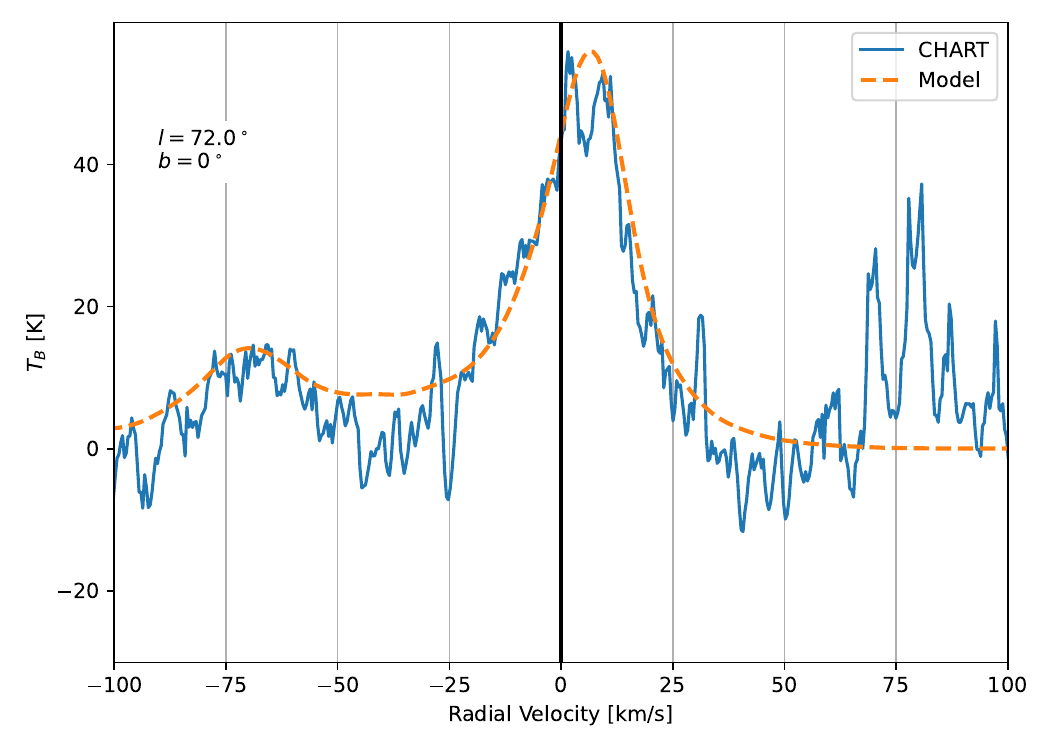}
    \captionof{figure}{Observed Brightness Temperature (TB) spectrum  as compared to the simulation, with a pointing of galactic coordinates $l=72^\circ$, $b=0^\circ$. This follows the $l=72^\circ$  line in figure \ref{fig:milkyway}. This data has been masked and filtered in the Fourier domain. The observed data peak width and center appear to be well matched to the model. There is a ripple across the data on the level of the high negative velocity clouds, so the match of the non-zero velocity peaks cannot be determined from this data. }
    \label{fig:phxcleaned}
\end{Figure}

The data was taken at galactic coordinates $l=72^\circ$ and $b=0^\circ$, following the $l=72^\circ$ line of sight in figure \ref{fig:milkyway}. 

For this data, a few extra analysis steps were taken to remove contaminants. 

Using a reference pointing with the horn covered, the data was masked for the contaminants that show up in both the control data and the sky data. We know sources that show up with the horn covered to be self or environmental interference rather than astronomical in nature, and can safely remove them from the data. We also applied a Fourier domain filter to remove any frequency dependent systematics. 

 The filtered and calibrated signal is seen in figure \ref{fig:phxcleaned}. The data appears well matched in width and peak to the simulation. The noise level of this data is high, so we do not interpret the negative velocity peak as a true match. There is a ripple across the band on the level of this peak, as seen in the bump at positive velocities of 75 km/s. However, the main lobe of our emission line seems to be an excellent fit and is well above the noise level of the ripple.

\section{Using CHART in the Classroom}\label{class}

At the time of this manuscript, CHART has been used in the classroom at the undergraduate and High School level.

Undergraduate students take responsibility for many of the development and testing parts of the project. This offers students an easy entry point to research, as the topics we teach with CHART (antenna design, signal processing, etc.) are essential to understanding research with larger scale experiments. To this front, we host summer undergraduate internships at Arizona State University (ASU) and Winona State University (WSU) that focus on CHART. We have also taught an entry level week-long workshop in Spring 2023, attended by 30 undergraduate students, as well as a Spring 2021 semester upper level course which registered 12 students at ASU.

At the High School level, CHART has been used at WSU for a week-long Summer workshop. The workshop brought four teachers, mostly from the areas surrounding Winona, to WSU and covered the introductory science, radio instrument construction and building, and observing and data analysis for the project. All participants were able to successfully detect galactic hydrogen, and left the workshop with the provided components for future use.

Quantitative details of the impacts of the project on students \& educators will be left to the work of a future study.

\section{Conclusion}
The CHART platform offers a classroom accessible method for students and educators to get involved in radio astronomy.  Here we have demonstrated detection and analysis of the 21-cm line from local neutral hydrogen using a cardboard horn, low cost electronics and freely available software. 

The base platform can be modified and extended by advanced users.
Future directions include further analysis to create galactic rotation curves demonstrating the necessity of dark matter and alternative radio targets at other bands.
With some component changes, the platform is easily extensible to other projects including solar observations, radio Jove (Jupiter), pulsars, and Cosmic Microwave Background (CMB) observations. Additionally, with multiple CHART systems, interferometry is possible. 

The documentation and construction of this project is designed to be easily accessible to secondary school level educators. 

CHART has been used at summer workshops for High School teachers, as well as by local teachers in the areas surrounding participating universities. We also engage undergraduates in the project through research involvement and coursework. We hope to expand project engagement even further as the documentation and science results grow.

\section*{Acknowledgements}

LMB acknowledges that this material is based upon work supported by a National Science Foundation Graduate Research Fellowship under Grant No. 2233001.

APB acknowledges support from an NSF Astronomy and Astrophysics Postdoctoral Fellowship under award AST-1701440, and a grant from the Mt. Cuba Astronomical Foundation.

APB and AW acknowledge support from NSF grant AST-2108348.

DCJ acknowledges NSF CAREER, grant \#2144995.

For the modeling data in this paper, we acknowledge the EU-HOU project and the Comenius grant.

\printbibliography

\end{multicols}

\end{document}